%% file: conf.tex
\documentclass[preprint,prd,tightenlines,superscriptaddress]{revtex4-1}

\usepackage{bm,amsmath,amsfonts,amssymb,graphicx,xspace,placeins,multirow}

\usepackage{graphicx} % Include figure files
\usepackage{dcolumn}  % Align table columns on decimal point
\usepackage{colordvi}
\usepackage{color}
\usepackage{epstopdf}
\usepackage{amssymb}
\usepackage{url}
\graphicspath{{ps}}
\usepackage{hyperref}
\usepackage{tabularx}
\usepackage{lineno}

\newcommand{\bdslnu}{\ensuremath{\Bzb \to D^{*+} \ell^{-} \nub_\ell}\xspace}

\newcommand{\bdlnu}{\ensuremath{\Bm \to D^{0} \ell^{-} \nub_\ell}\xspace}
\newcommand{\bdenu}{\ensuremath{\Bm \to D^{0} e^{-} \nub_e}\xspace}
\newcommand{\bdmunu}{\ensuremath{\Bm \to D^{0} \mu^{-} \nub_\mu}\xspace}

\newcommand{\ifb}{\ensuremath{{\rm fb}^{-1}}\xspace}

\newcommand{\BR}{{\ensuremath{\cal B}}}

\newcommand{\cosby}{\ensuremath{\cos\theta_{BY}}}

\newcommand{\NBB}{\ensuremath{N_{B \bar B}=  \left(68.21 \pm 0.06_{\text{stat}}\pm 0.75_{\text{sys}}\right)\times 10^6}}

\input ./belle2sym.tex

\begin{document}
% \linenumbers

%place for definitions and newcommands
\def\belletwo {\it {Belle II}}

\vspace*{-3\baselineskip}
\resizebox{!}{3cm}{\includegraphics{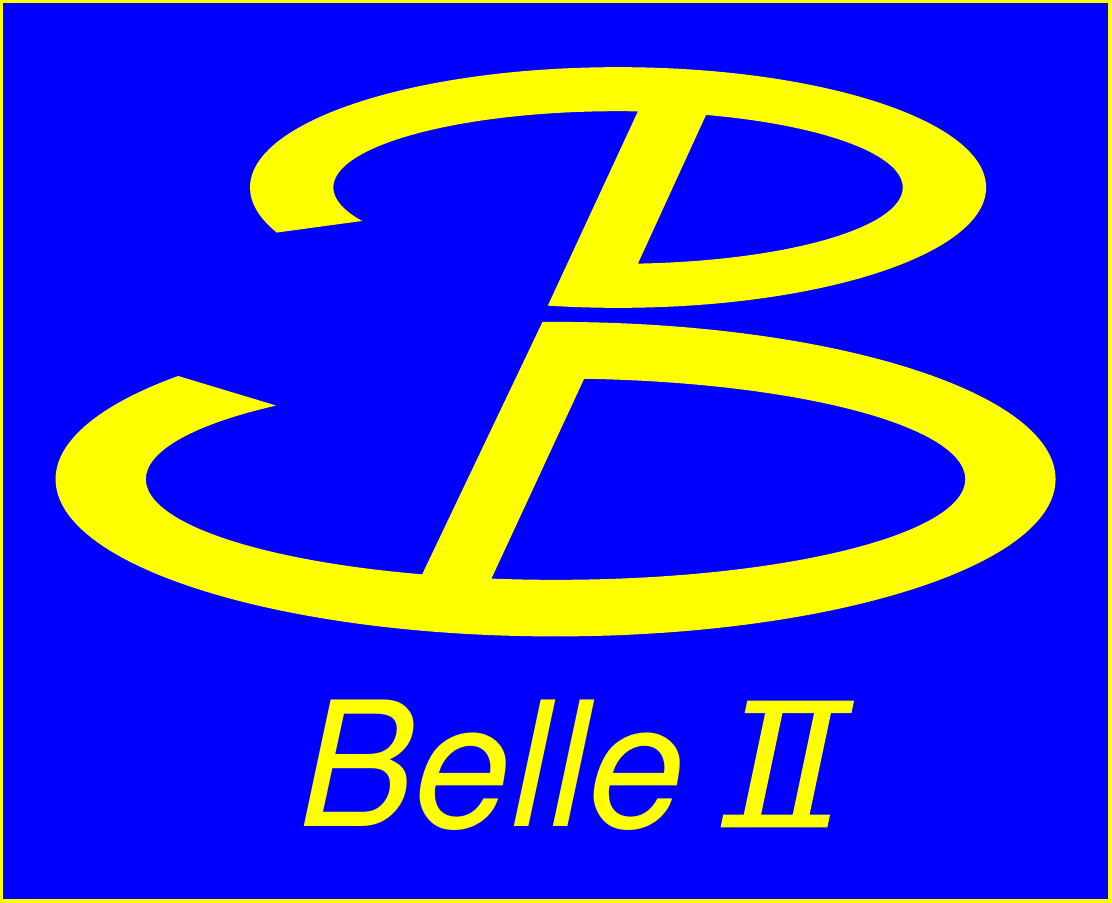}}

\vspace*{-5\baselineskip}
    \begin{flushright}
BELLE2-CONF-PH-2021-011 \\
\today
\end{flushright}

\title { \quad\\[0.5cm] Measurement of the $B^-\to D^0 \ell^- \bar\nu_\ell$ Branching Fraction in 62.8~fb$^{-1}$ of Belle II data}

% \collaboration{The Belle II Collaboration}
\input{authors-conf2020}

\begin{abstract}
We report a measurement of the branching fraction of the semileptonic decay \bdlnu (and its charge conjugate) using \mbox{$62.8~\text{fb}^{-1}$} of $\Upsilon(4S)\to B\bar B$~data recorded by the Belle II experiment at the SuperKEKB asymmetric-energy $e^+ e^-$ collider. The neutral charm meson is searched for in the decay mode $D^0\to K^-\pi^+$ and combined with a properly charged identified lepton (electron or muon) to reconstruct this decay. No reconstruction of the second $B$~meson in the $\Upsilon(4S)$~event is performed. We obtain ${\cal B}(B^-\to D^0 \ell^-\bar\nu_\ell) =  (2.29\pm 0.05_\mathrm{stat}\pm 0.08_\mathrm{syst})\%$, in agreement with the world average of this decay. We also determine the ratio of the electron to muon branching fractions to be $R(e/\mu) = 1.04 \pm 0.05_\mathrm{stat} \pm 0.03_\mathrm{syst}$ and observe no deviation from lepton universality.

\end{abstract}


\maketitle

{\renewcommand{\thefootnote}{\fnsymbol{footnote}}}
\setcounter{footnote}{0}

\section{Introduction}

The magnitude of the Cabibbo-Kobayashi-Maskawa (CKM)~\cite{Cabibbo:1963yz,Kobayashi:1973fv} matrix element $|V_{cb}|$ squared determines the transition rate of $b$- into $c$-quarks and the precise knowledge of this fundamental parameter of the Standard Model (SM)~\cite{Pich:2007vu} is crucial for the ongoing precision-$B$-physics programme at the Belle II experiment and elsewhere. The CKM element $|V_{cb}|$ is measured from semileptonic $B$~meson decays $B\to X_c\ell\nu$, where $X_c$ is a hadronic system with charm, $\ell$ is a light charged lepton (electron or muon) and $\nu$ is the associated neutrino. These determinations can be $\emph{inclusive}$, {\it i.e.}, sensitive to all $X_c\ell\nu$ final states within a given region of phase space, or $\emph{exclusive}$, {\it i.e.}, based only on a single $b\to c$ semileptonic mode such as $B\to D^*\ell\nu$ or $B\to D\ell\nu$. Pursuing both approaches is important as the two avenues involve different theoretical and experimental uncertainties and consistency between both is a powerful cross-check of our understanding. However, inclusive and exclusive measurements of $|V_{cb}|$ are at odds for many years now, an issue which is often referred to as the \emph{inclusive vs.\ exclusive problem}~\cite{Amhis:2019ckw}.

In this paper we describe the measurement of the branching fraction of the decay \bdlnu~\cite{cc}, a mode which is expected to yield a precise determination of the CKM element $|V_{cb}|$ from the Belle II data. Neutral $D$~mesons are searched for in the decay mode $D^0\to K^-\pi^+$ and combined with an identified lepton (electron or muon) of the same charge as the kaon to reconstruct this decay. To maximize the statistical power of the early Belle II data, this analysis is untagged, {\it i.e.}, we do not place any constraint on the second $B$~meson in the $\Upsilon(4S)$ event. The paper is organized as follows: Sect.~\ref{sec:dataset} describes the real data and simulated data sets used throughout this analysis. The experimental procedure is described in Sect.~\ref{sec:procedure}. Finally, Sect.~\ref{sec:results} contains the results of this analysis.

\section{The Belle~II detector and data sample} \label{sec:dataset}

The Belle~II detector~\cite{Abe:2010sj, ref:b2tip} operates at the SuperKEKB asymmetric-energy  electron-positron collider~\cite{superkekb}, located at the KEK laboratory in Tsukuba, Japan. The detector consists of several nested detector subsystems arranged around the beam pipe in a cylindrical geometry. The innermost subsystem is the vertex detector, which includes two layers of silicon pixel detectors and four outer layers of silicon strip detectors. Currently, the second pixel layer is installed in only a small part of the solid angle, while the remaining vertex detector layers are fully installed. Most of the tracking volume consists of a helium and ethane-based small-cell drift chamber (CDC). Outside the drift chamber, a Cherenkov-light imaging and time-of-propagation detector provides charged-particle identification in the barrel region. In the forward endcap, this function is provided by a proximity-focusing, ring-imaging Cherenkov detector with an aerogel radiator. Further out is the ECL electromagnetic calorimeter, consisting of a barrel and two endcap sections made of CsI(Tl) crystals. A uniform 1.5~T magnetic field is provided by a superconducting solenoid situated outside the calorimeter. Multiple layers of scintillators and resistive plate chambers, located between the magnetic flux-return iron plates, constitute the $K_L$ and muon identification system (KLM).

The data used in this analysis were collected in the years 2019 and 2020 at a center-of-mass (c.m.) energy of 10.58~GeV, corresponding to the mass of the $\Upsilon$(4S) resonance.
%The energies of the electron and positron beams are $7\gev$ and $4\gev$, respectively, resulting in a boost of $\beta\gamma = 0.28$ of the CM frame relative to the lab frame.
This data set corresponds to an integrated luminosity of 62.8~fb$^{-1}$ and contains \NBB\ $\Upsilon(4S)\to B\bar B$ events as determined from a fit to event-shape variables~\cite{BELLE2-NOTE-PL-2020-006}.

Different samples of Monte Carlo (MC) simulated events are used throughout this analysis. These include a sample of $\Upsilon(4S)\to B\bar B$~events in which $B$~mesons decay generically, generated with EvtGen~\cite{Lange:2001uf} and a sample of continuum $e^+e^-\to q\bar q$~events ($q = u, d, s, c$) simulated with KKMC~\cite{Ward:2002qq}, interfaced with PYTHIA~\cite{Sjostrand:2007gs}. Full detector simulation based on GEANT4~\cite{ref:geant4} is applied to MC~events. The Monte Carlo samples used in this analysis correspond to an integrated luminosity of 300~\ifb. The lepton reconstruction efficiencies and the hadron misidentification rates in simulation are adjusted to match the real performance of the Belle II lepton identification system.

The data samples are processed using the Belle II software framework basf2~\cite{Kuhr:2018lps}.

Prior to physics analysis, charged particle trajectories are reconstructed in the vertex detector and the central drift chamber~\cite{Bertacchi:2020eez}. Photons are reconstructed from ECL clusters unmatched to charged particle tracks. Hadronic events are selected by requiring at least three charged particles, a visible energy above 4~GeV, and a ratio $R_2$ of the second to the zeroth Fox-Wolfram moments below 0.3~\cite{Fox:1978vu}.

\section{Experimental procedure} \label{sec:procedure}

\subsection{Reconstruction}

We require charged particle tracks to originate from the interaction point (IP): The distance of closest approach between each track and the interaction point must be less than 2~cm along the $z$ direction (parallel to the beams) and less than 0.5~cm in the transverse $r-\phi$ plane. We further require charged particles to be within angular acceptance of the central drift chamber and to have associated CDC hits.

Charged leptons (electron or muons) are required to have a c.m.\ momentum greater than 0.6~GeV. Electrons are identified based on their energy and shower shape in the ECL calorimeter. Muons are found based on the information of the instrumented return yoke KLM. We attempt to recover bremsstahlung photons radiated from an electron track by searching within a cone around the lepton direction. If such photons, with an energy between 50~MeV and 150~MeV, are found they are added to the electron candidate to correct the 4-momentum.

Neutral $D$~meson candidates are searched for in the decay mode to $K^-\pi^+$, $D^0\to K^-\pi^+$. $D^0$~candidates are accepted within a $K\pi$~invariant mass window from 1.857~GeV to 1.872~GeV.

Candidates for the decay \bdlnu are obtained by combining an appropriately charged lepton with a neutral $D$~candidate. The mass of the $Y = D^0\ell$ system is required to exceed 3.15 GeV. For each $B$~candidate, we calculate the angle between the $Y$ and the $B$ meson in the c.m.\ frame of the collision,
\begin{equation}
\cos\theta_{BY} = {2\, E_B^* E_Y^* - m_B^2 - m_Y^2 \over 2 |p_B^*||p_Y^*|},
\end{equation}
where $E_Y^*$, $|p_Y^*|$, and $m_Y$ are the c.m.\ energy, momentum, and invariant mass, respectively, of the $D^0\ell$ system, $m_B$ is the nominal $B$ mass~\cite{pdg:2020}, and $E_B^*$, $|p_B^*|$ are the c.m.\ energy and momentum, respectively, of the $B$. The latter are inferred from the beam 4-momenta. For correctly reconstructed \bdlnu~candidates, the value of $\cos\theta_{BY}$ ranges within the interval $[-1,1]$. However, due to the finite beam-energy spread, final-state radiation, and detector resolution, the $\cos\theta_{BY}$ distributions of signal events are smeared beyond this range. For background candidates, values outside of the $[-1,1]$ interval are allowed. In the rest of the analysis, we retain $B$~candidates with a value of $\cos\theta_{BY}$ ranging between $-4$ and 4.

To reduce the sizeable background of $B^0\to D^{*-}(\bar D^0\pi^-)\ell^+\nu_\ell$ and $B^+\to\bar D^{*0}(\bar D^0\pi^0)\ell^+\nu_\ell$~decays, an active veto is applied. For $B^0\to D^{*-}(\bar D^0\pi^-)\ell^+\nu_\ell$, this is done by combining a slow ($p<0.35$~GeV) pion of correct charge with the $D^0$ of a $B^+\to\bar D^0\ell^+\nu_\ell$ candidate. If, for any slow pion candidate in the event, the mass difference $\Delta M=M(D^*)-M(D)$ is found to be in the interval $[0.144,0.148]$~GeV, the $B^+$~candidate is rejected. For $B^+\to\bar D^{*0}(\bar D^0\pi^0)\ell^+\nu_\ell$~decays, we combine the $D^0$ with a neutral pion candidate and reject the \bdlnu candidate, if $\Delta M$ is in the interval $[0.141,0.145]$~GeV and the opening angle between $D^0$ and $\pi^0$ is below 17~degrees. We reconstruct neutral pions from $\pi^0 \rightarrow \gamma \gamma$ and require different energies of the photon daughters depending on the region of the detector the photon signature originated from. We require $E > 0.080$~GeV for the forward end-cap, $E > 0.030$~GeV for the barrel region and $E > 0.060$~GeV for the backward end-cap. The $\pi^0$ mass is required to be in the interval $[0.120,0.145]$~GeV.

Fig.~\ref{fig:prefit} shows the $\cos\theta_{BY}$ distributions of \bdenu and \bdmunu candidates after applying the selections described in this section.
\begin{figure}
    \centering
    \includegraphics[width=.48\columnwidth]{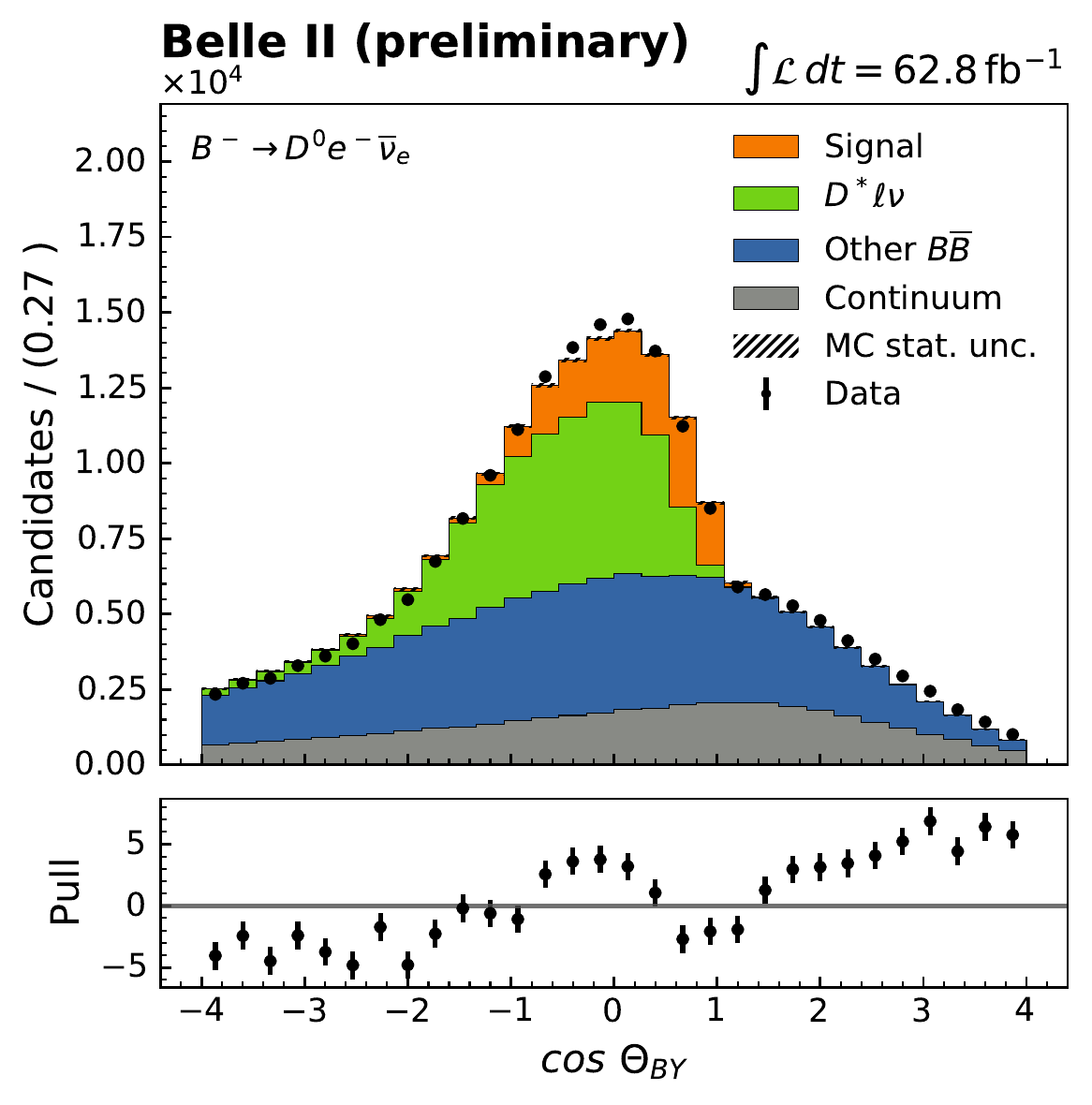}
    \includegraphics[width=.47\columnwidth]{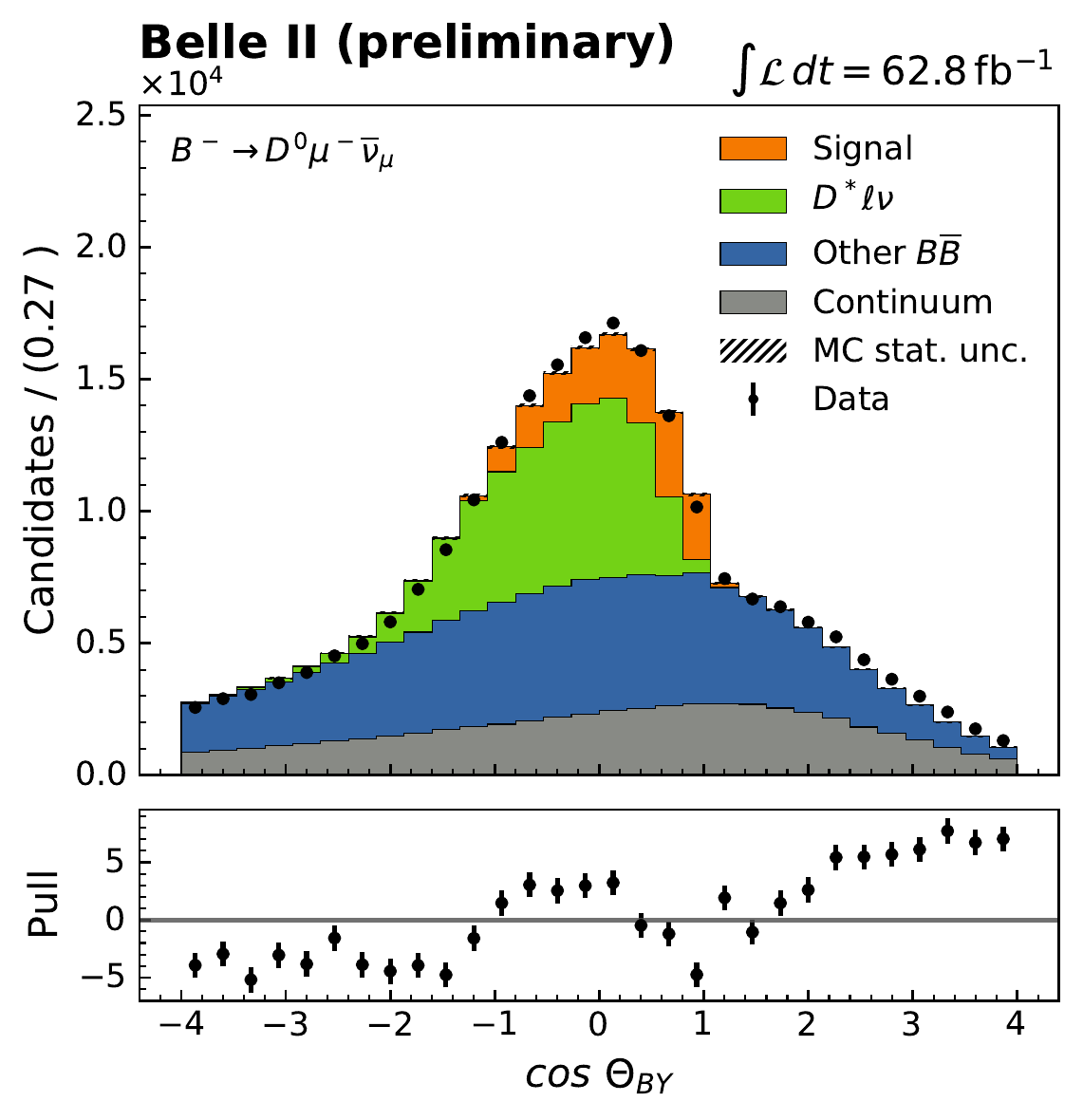}
    \caption{$\cos\theta_{BY}$~distributions for selected \bdenu (left) and \bdmunu candidates (right). The stacked histograms are MC simulated events scaled to the real data luminosity of 62.8~fb$^{-1}$. The real data is shown by points with error bars.}
    \label{fig:prefit}
\end{figure}

\subsection{Signal extraction}

To extract the amount of signal in the selected sample, we perform separate fits to the $\cos\theta_{BY}$ distributions of \bdenu and \bdmunu candidates. We use a maximum likelihood technique using Poisson statistics of both real and MC simulated data~\cite{Barlow:1993dm}. The MC shape of the signal, $D^*$~downfeed, background from $B\bar B$~events and continuum background distributions is kept, while the respective normalizations are free parameters in both fits.

\begin{figure}
    \centering
    \includegraphics[width=.48\columnwidth]{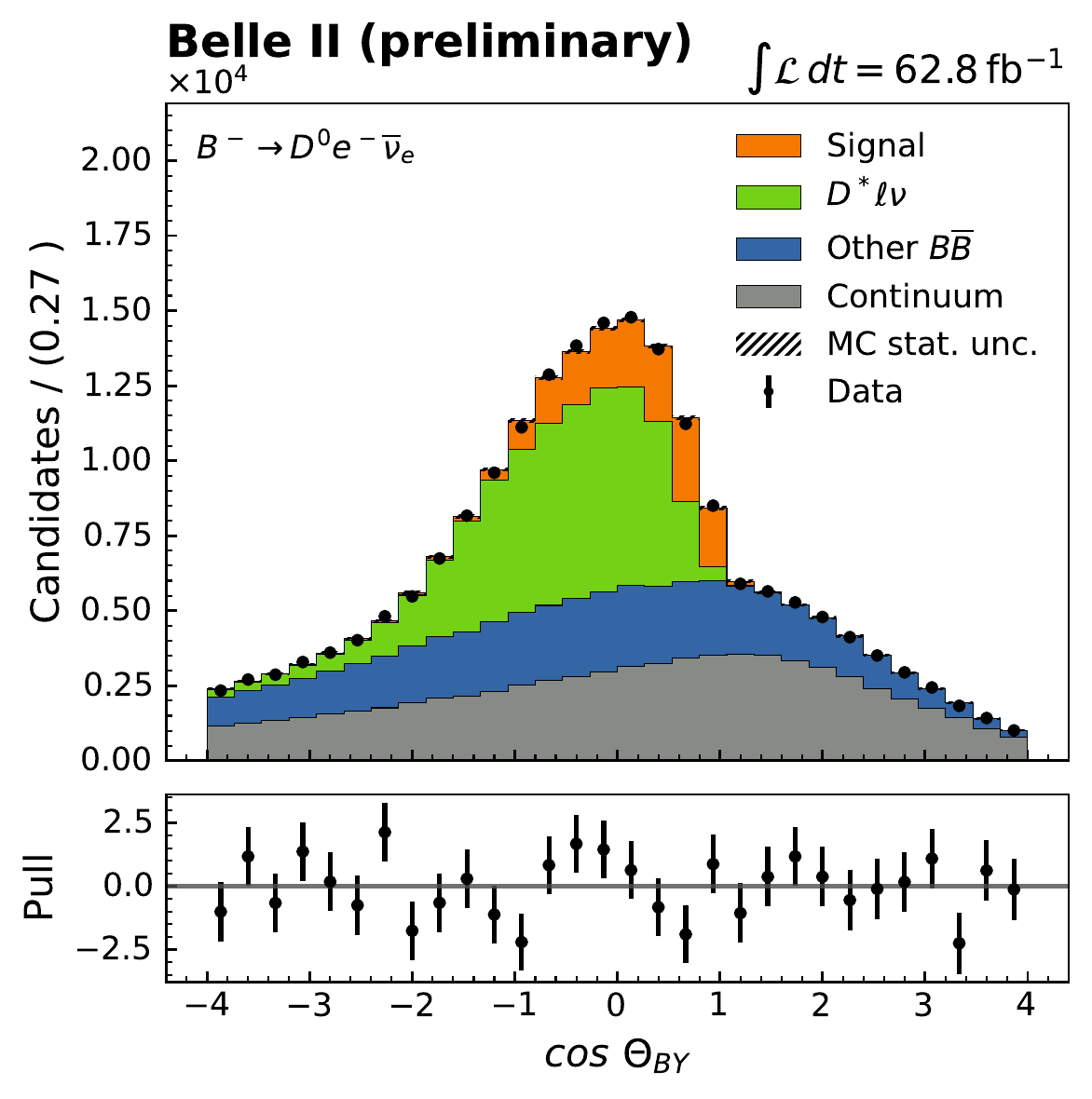}
    \includegraphics[width=.47\columnwidth]{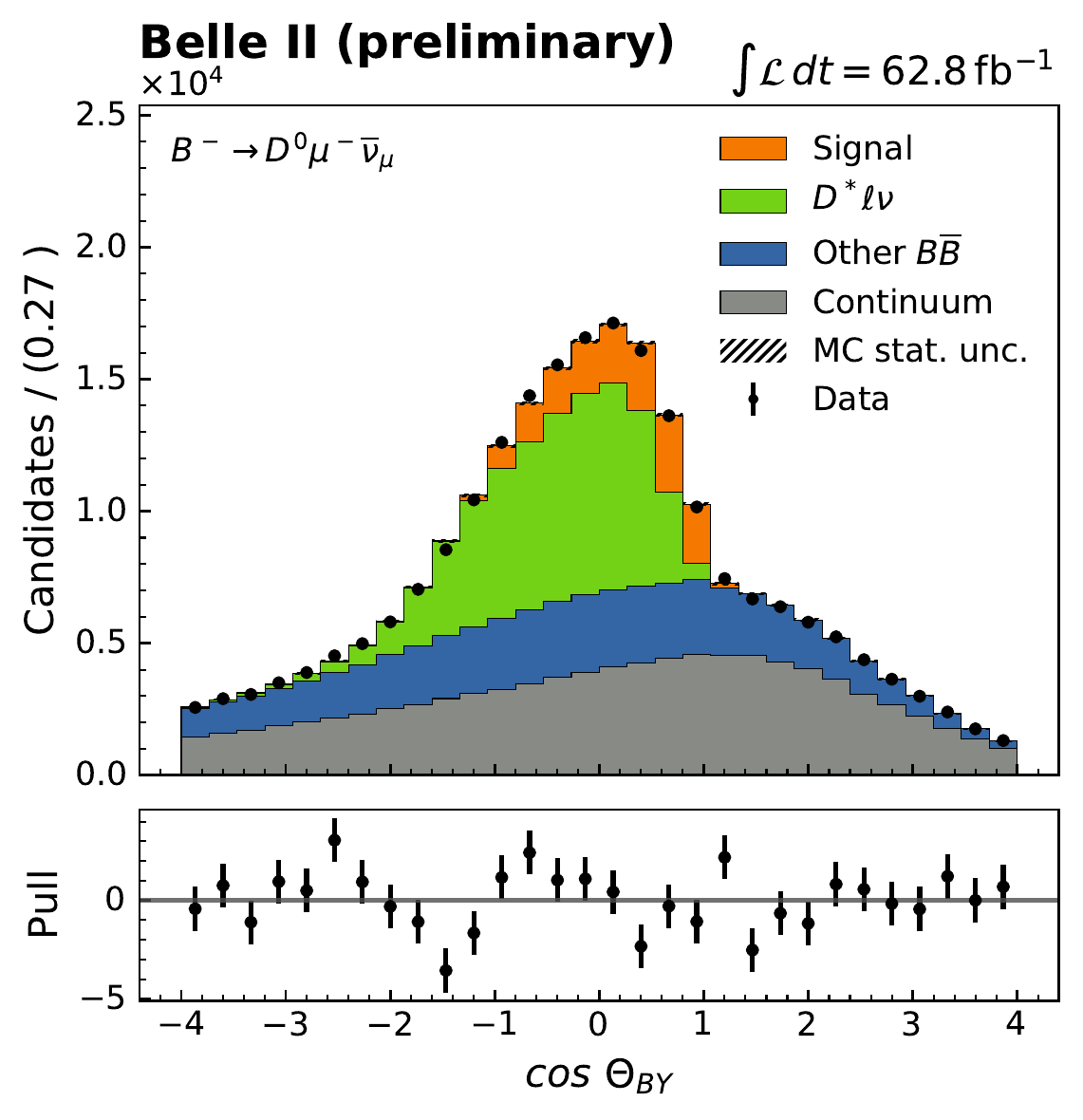}
    \caption{Result of the fit to the \bdenu (left) and \bdmunu samples (right). The stacked histograms are MC simulated events scaled to match the result of the fit. The real data is shown by points with error bars.}
    \label{fig:postfit}
\end{figure}
The fit results are shown in Table~\ref{tab:results}. We find $19,543\pm 628$ ($18,869\pm 636$) events in the electron (muon) channel. In Fig.~\ref{fig:postfit} the stacked MC components are scaled according to the fit results and the collision data is shown by points with error bars. 

\begin{table}
    \centering
    \begin{tabular}{c|c|c|c|c}
        \hline \hline
        \multicolumn{1}{c}{} & \multicolumn{4}{|c}{Fitted Yields} \\ \hline
        Channel & Signal & $D^*$ & Other $B\overline{B}$ & Continuum\\
        \hline
        \bdenu & $19543\pm 648$&  $65502 \pm 960$& $59233 \pm 2450$ &$79697 \pm 1970$  \\
        \bdmunu & $18869 \pm 636$& $67595 \pm 843$& $64899 \pm 2101$& $102308 \pm 1808$\\ \hline \hline
    \end{tabular}
    \caption{The fitted yield for each MC component determined from a maximum likelihood fit in \cosby. The uncertainties are statistical only.} \label{tab:results}
\end{table}

\section{Results and systematic uncertainty} \label{sec:results}

\subsection{\bdlnu branching fraction}

The fit result quoted in the previous section can be converted into a measurement of the \bdlnu branching ratio by using
\begin{equation}
    N_\mathrm{sig}=2\times N_{B\bar B}\times f_{+-}\times {\cal B}(\bdlnu)\times {\cal B}(D^0\to K^-\pi^+)\times\epsilon~,
\end{equation}
where $N_{B\bar B}$ is the number of $\Upsilon(4S)$~events in the sample, $f_{+-}$ is the $B^+$~production fraction at the $\Upsilon(4S)$~\cite{pdg:2020}, ${\cal B}(D^0\to K^-\pi^+)$ is the $D^0$~subdecay branching fraction~\cite{pdg:2020} and $\epsilon$ is the overall selection criteria efficiency of this analysis as determined from MC simulation.  The results obtained in the two samples are collected in Table~\ref{tab:results_br}.

\begin{table}
    \centering
    \begin{tabular}{c|c|c}
        \hline \hline
        Channel & Efficiency [\%] & Branching fraction [\%]\\
        \hline
        \bdenu & 30.12 & $2.34\pm 0.08$\\
        \bdmunu & 30.36 & $2.24\pm 0.08$\\ \hline \hline
    \end{tabular}
    \caption{Branching fractions of \bdlnu determined in the electron and muon samples. The uncertainties are statistical only.} \label{tab:results_br}
\end{table}

%%%%%%%%%%%%%%%%%%%%%%%%%
\subsection {Systematic uncertainty} \label{sec:syst}

The relative systematic uncertainties affecting the \bdlnu branching fraction measurement are listed in Table~\ref{tab:syst}. We assume no correlation among the individual sources of uncertainty and sum them in quadrature to obtain the total systematic uncertainty. The  methods used for obtaining these uncertainties are detailed below.

\begin{table}
    \centering
    \begin{tabular}{l|c|c}
        \hline \hline
         & \multicolumn{2}{c}{Relative uncertainty [\%]} \\
        \cline{2-3}
        Source &$B^-\to D^0 e^- \overline{\nu}_{e}$&$B^-\to D^0 \mu^- \overline{\nu}_{\mu}$ \\
        \hline
        $N_{B^{\pm}}$ & 1.61 & 1.61\\
        ${\cal B}(D^0\to K^-\pi^+)$ & 0.78 & 0.78\\
        Tracking & 2.07 & 2.07\\
        Lepton identification & 1.41 & 2.38\\
        MC efficiency (statistical) & 0.09 & 0.09\\
        $D\ell\nu$ form factor & 0.15 & 0.15\\
        $D^*\ell\nu$ form factor & 0.44 & 0.44\\
        Continuum shape & 0.37 & 0.37\\
        \hline
        Sum & 3.14 & 3.68\\
        \hline \hline
    \end{tabular}
    \caption{Relative systematic uncertainty on the measurement of the \bdlnu branching fraction in the two samples.} \label{tab:syst}
\end{table}

To correct for mismodelling of the lepton-identification in the MC compared to collision events, we apply momentum-and polar-angle-dependent corrections.
In independent studies of $J/\psi\to \ell^+\ell^-$ and $K_S \to \pi^+ \pi^-$ decays, correction factors are obtained for the reconstruction efficiency of leptons, and the mis-identification of hadrons as leptons. 
Due to limited sample size in the control samples, the lepton-identification correction factors are associated with statistical and systematic uncertainties. By resampling the correction factors with Gaussian distributions, while accounting for systematic correlations, we generate 500 sets of correction values. The 500 sets are used to estimate the systematic uncertainty on $N_\mathrm{sig}$ caused by lepton-identification.

The uncertainty on the branching fraction of the hadronic decay mode  $\BR(\Dz \to \Km \pip)$ = $(3.950\pm0.031)\%$~\cite{Zyla:2020} enters the result of the signal yield as a systematic uncertainty.

The number of charged $B^{\pm}$ mesons in the data sample is calculated as
\begin{equation}
   N_{B^{\pm}} = 2 \times N_{B\bar B} \times f_{+-}~\,
\end{equation}
with \NBB~and
\begin{equation}
  f_{+-} = \frac{\Gamma\left(\Upsilon\left(4S\right) \to B^+ B^-\right)}{\Gamma\left(\Upsilon\left(4S\right)\right)_\mathrm{tot} } = 0.514 \pm 0.006.
\end{equation}
The uncertainties on $f_{+-}$ and $N_{B \bar B}$ are added in quadrature to estimate the impact on the measured branching fraction.

We account for the effect of finite MC sample sizes on the selection efficiency $\epsilon$ with the binomial standard error.

A $e^+e^- \to \tau^+ \tau^-$ performance study measures discrepancies in the track finding efficiency between MC and collision data. In accordance with the performance study, a relative systematic uncertainty of $0.69~\%$ is assigned for each of the three charged final state tracks to account for the track efficiency discrepancy.

The form factors describe the dependency of the decay rate on the kinematic variable $w = v_B \cdot v_{D^{(*)}}$. The form factors impact on the shape of signal and $D^*$ components has to be taken into account. We separately vary the form factor parameters of the decays \bdlnu and \bdslnu in the parameterization of Caprini, Lellouch and Neubert (CLN)~\cite{Caprini:1997mu} by $1~\sigma$ around their central values~\cite{HFLAV:2019otj} to estimate the corresponding systematic uncertainty. The form factor uncertainty quoted in Table~\ref{tab:syst} corresponds to the quadratic sum of these individual variations.

Finally, the discrepancies between data and MC in the sidebands of the pre-fit \cosby~distributions in Fig.~\ref{fig:prefit} are partly explained by mismodelling of the continuum MC. We estimate the effect of this mismodelling on the measured branching fractions by reweighing the continuum MC using collision data recorded below the $\Upsilon(4S)$.

%%%%%%%%%%%%%%
\section{Summary}

We have measured the branching fraction of the decay \bdlnu in 62.8~fb$^{-1}$ of Belle~II data. The results in the electron and muon samples are
\begin{eqnarray}
  {\cal B}(\bdenu) & = & (2.34\pm 0.08_\mathrm{stat}\pm 0.07_\mathrm{syst})\%~,\\
  {\cal B}(\bdmunu) & = & (2.24\pm 0.08_\mathrm{stat}\pm 0.08_\mathrm{syst})\%~,
\end{eqnarray}
where the first error is statistical and the second systematic.

The weighted mean of both modes yields to this combined value of the branching fraction
\begin{equation}
  {\cal B}(\bdlnu) =  (2.29\pm 0.05_\mathrm{stat}\pm 0.08_\mathrm{syst})\%~,
\end{equation}
in agreement with the world average value of $(2.35\pm 0.03_\mathrm{stat}\pm 0.09_\mathrm{syst})\%$~\cite{HFLAV:2019otj}. For the ratio between the $e$ and $\mu$ channels, the uncertainties related to $N_{B^{\pm}}$ and ${\cal B}(D^0\to K^-\pi^+)$ cancel and we obtain
\begin{equation}
    R(e/\mu) = \frac{{\cal B}(\bdenu)}{{\cal B}(\bdmunu)}= 1.04\pm 0.05_\mathrm{stat}\pm 0.03_\mathrm{syst}~.
\end{equation}

\section{Acknowledgments}
\input{acknowledgements}

\bibliography{conf.bib}

\end{document}

%% file: authors-conf2020.tex
%%% Paper:    (2021 conference papers)
%%% Journal:  (2021 conferences)
%%% Updates:
%%% July 21, 2021 - new iteration
%%% ====================================================================
%%% Use \input{authors-conf2021} to insert this material into your latex file.
%%%\newcommand{\instSinica}{Academia Sinica, Taipei 11529, Taiwan}
\newcommand{\instCPPM}{Aix Marseille Universit\'{e}, CNRS/IN2P3, CPPM, 13288 Marseille, France}
\newcommand{\instYerevan}{Alikhanyan National Science Laboratory, Yerevan 0036, Armenia}
\newcommand{\instBeihang}{Beihang University, Beijing 100191, China}
\newcommand{\instBNL}{Brookhaven National Laboratory, Upton, New York 11973, U.S.A.}
\newcommand{\instBINP}{Budker Institute of Nuclear Physics SB RAS, Novosibirsk 630090, Russian Federation}
\newcommand{\instCMU}{Carnegie Mellon University, Pittsburgh, Pennsylvania 15213, U.S.A.}
\newcommand{\instCinvestavIPN}{Centro de Investigacion y de Estudios Avanzados del Instituto Politecnico Nacional, Mexico City 07360, Mexico}
\newcommand{\instPrague}{Faculty of Mathematics and Physics, Charles University, 121 16 Prague, Czech Republic}
\newcommand{\instChiangMai}{Chiang Mai University, Chiang Mai 50202, Thailand}
\newcommand{\instChiba}{Chiba University, Chiba 263-8522, Japan}
\newcommand{\instChonnam}{Chonnam National University, Gwangju 61186, South Korea}
\newcommand{\instConacyt}{Consejo Nacional de Ciencia y Tecnolog\'{\i}a, Mexico City 03940, Mexico}
\newcommand{\instDESY}{Deutsches Elektronen--Synchrotron, 22607 Hamburg, Germany}
\newcommand{\instDuke}{Duke University, Durham, North Carolina 27708, U.S.A.}
\newcommand{\instITAR}{Institute of Theoretical and Applied Research (ITAR), Duy Tan University, Hanoi 100000, Vietnam}
\newcommand{\instRomaENEA}{ENEA Casaccia, I-00123 Roma, Italy}
\newcommand{\instFuJen}{Department of Physics, Fu Jen Catholic University, Taipei 24205, Taiwan}
\newcommand{\instFudan}{Key Laboratory of Nuclear Physics and Ion-beam Application (MOE) and Institute of Modern Physics, Fudan University, Shanghai 200443, China}
\newcommand{\instGoettingen}{II. Physikalisches Institut, Georg-August-Universit\"{a}t G\"{o}ttingen, 37073 G\"{o}ttingen, Germany}
\newcommand{\instGifu}{Gifu University, Gifu 501-1193, Japan}
\newcommand{\instSOKENDAI}{The Graduate University for Advanced Studies (SOKENDAI), Hayama 240-0193, Japan}
\newcommand{\instGyeongsang}{Gyeongsang National University, Jinju 52828, South Korea}
\newcommand{\instHanyang}{Department of Physics and Institute of Natural Sciences, Hanyang University, Seoul 04763, South Korea}
\newcommand{\instKEK}{High Energy Accelerator Research Organization (KEK), Tsukuba 305-0801, Japan}
\newcommand{\instJPARC}{J-PARC Branch, KEK Theory Center, High Energy Accelerator Research Organization (KEK), Tsukuba 305-0801, Japan}
\newcommand{\instHiroshima}{Hiroshima University, Higashi-Hiroshima, Hiroshima 739-8530, Japan}
\newcommand{\instFrascati}{INFN Laboratori Nazionali di Frascati, I-00044 Frascati, Italy}
\newcommand{\instNapoliINFN}{INFN Sezione di Napoli, I-80126 Napoli, Italy}
\newcommand{\instPadovaINFN}{INFN Sezione di Padova, I-35131 Padova, Italy}
\newcommand{\instPerugiaINFN}{INFN Sezione di Perugia, I-06123 Perugia, Italy}
\newcommand{\instPisaINFN}{INFN Sezione di Pisa, I-56127 Pisa, Italy}
\newcommand{\instRomaINFN}{INFN Sezione di Roma, I-00185 Roma, Italy}
\newcommand{\instRomaTreINFN}{INFN Sezione di Roma Tre, I-00146 Roma, Italy}
\newcommand{\instTorinoINFN}{INFN Sezione di Torino, I-10125 Torino, Italy}
\newcommand{\instTriesteINFN}{INFN Sezione di Trieste, I-34127 Trieste, Italy}
\newcommand{\instIISER}{Indian Institute of Science Education and Research Mohali, SAS Nagar, 140306, India}
\newcommand{\instIITBhubaneswar}{Indian Institute of Technology Bhubaneswar, Satya Nagar 751007, India}
\newcommand{\instIITGuwahati}{Indian Institute of Technology Guwahati, Assam 781039, India}
\newcommand{\instIITHyderabad}{Indian Institute of Technology Hyderabad, Telangana 502285, India}
\newcommand{\instIITMadras}{Indian Institute of Technology Madras, Chennai 600036, India}
\newcommand{\instIndiana}{Indiana University, Bloomington, Indiana 47408, U.S.A.}
\newcommand{\instIHEPRussia}{Institute for High Energy Physics, Protvino 142281, Russian Federation}
\newcommand{\instHEPHYVienna}{Institute of High Energy Physics, Vienna 1050, Austria}
\newcommand{\instIHEPChina}{Institute of High Energy Physics, Chinese Academy of Sciences, Beijing 100049, China}
\newcommand{\instIPP}{Institute of Particle Physics (Canada), Victoria, British Columbia V8W 2Y2, Canada}
\newcommand{\instIOP}{Institute of Physics, Vietnam Academy of Science and Technology (VAST), Hanoi, Vietnam}
\newcommand{\instIFIC}{Instituto de Fisica Corpuscular, Paterna 46980, Spain}
\newcommand{\instISU}{Iowa State University, Ames, Iowa 50011, U.S.A.}
\newcommand{\instJAEA}{Advanced Science Research Center, Japan Atomic Energy Agency, Naka 319-1195, Japan}
\newcommand{\instMainz}{Institut f\"{u}r Kernphysik, Johannes Gutenberg-Universit\"{a}t Mainz, D-55099 Mainz, Germany}
\newcommand{\instGiessen}{Justus-Liebig-Universit\"{a}t Gie\ss{}en, 35392 Gie\ss{}en, Germany}
\newcommand{\instKarlsruhe}{Institut f\"{u}r Experimentelle Teilchenphysik, Karlsruher Institut f\"{u}r Technologie, 76131 Karlsruhe, Germany}
\newcommand{\instKitasato}{Kitasato University, Sagamihara 252-0373, Japan}
\newcommand{\instKISTI}{Korea Institute of Science and Technology Information, Daejeon 34141, South Korea}
\newcommand{\instKoreaUnivKU}{Korea University, Seoul 02841, South Korea}
\newcommand{\instKSU}{Kyoto Sangyo University, Kyoto 603-8555, Japan}
\newcommand{\instKyungpook}{Kyungpook National University, Daegu 41566, South Korea}
\newcommand{\instLPI}{P.N. Lebedev Physical Institute of the Russian Academy of Sciences, Moscow 119991, Russian Federation}
\newcommand{\instLNNU}{Liaoning Normal University, Dalian 116029, China}
\newcommand{\instLMU}{Ludwig Maximilians University, 80539 Munich, Germany}
\newcommand{\instLuther}{Luther College, Decorah, Iowa 52101, U.S.A.}
\newcommand{\instMNITJaipur}{Malaviya National Institute of Technology Jaipur, Jaipur 302017, India}
\newcommand{\instMPP}{Max-Planck-Institut f\"{u}r Physik, 80805 M\"{u}nchen, Germany}
\newcommand{\instMPGHLL}{Semiconductor Laboratory of the Max Planck Society, 81739 M\"{u}nchen, Germany}
\newcommand{\instMcGill}{McGill University, Montr\'{e}al, Qu\'{e}bec, H3A 2T8, Canada}
\newcommand{\instMEPhI}{Moscow Physical Engineering Institute, Moscow 115409, Russian Federation}
\newcommand{\instNagoya}{Graduate School of Science, Nagoya University, Nagoya 464-8602, Japan}
\newcommand{\instNagoyaIAR}{Institute for Advanced Research, Nagoya University, Nagoya 464-8602, Japan}
\newcommand{\instNagoyaKMI}{Kobayashi-Maskawa Institute, Nagoya University, Nagoya 464-8602, Japan}
\newcommand{\instNaraWu}{Nara Women's University, Nara 630-8506, Japan}
\newcommand{\instHSE}{National Research University Higher School of Economics, Moscow 101000, Russian Federation}
\newcommand{\instNTUTaiwan}{Department of Physics, National Taiwan University, Taipei 10617, Taiwan}
\newcommand{\instNUUTaiwan}{National United University, Miao Li 36003, Taiwan}
\newcommand{\instKrakow}{H. Niewodniczanski Institute of Nuclear Physics, Krakow 31-342, Poland}
\newcommand{\instNiigata}{Niigata University, Niigata 950-2181, Japan}
\newcommand{\instNSU}{Novosibirsk State University, Novosibirsk 630090, Russian Federation}
\newcommand{\instOkinawa}{Okinawa Institute of Science and Technology, Okinawa 904-0495, Japan}
\newcommand{\instOsakaCity}{Osaka City University, Osaka 558-8585, Japan}
\newcommand{\instRCNP}{Research Center for Nuclear Physics, Osaka University, Osaka 567-0047, Japan}
\newcommand{\instPNNL}{Pacific Northwest National Laboratory, Richland, Washington 99352, U.S.A.}
\newcommand{\instPanjab}{Panjab University, Chandigarh 160014, India}
\newcommand{\instPanjabPAU}{Punjab Agricultural University, Ludhiana 141004, India}
\newcommand{\instRIKENMSL}{Meson Science Laboratory, Cluster for Pioneering Research, RIKEN, Saitama 351-0198, Japan}
\newcommand{\instXavier}{St. Francis Xavier University, Antigonish, Nova Scotia, B2G 2W5, Canada}
\newcommand{\instSeoul}{Seoul National University, Seoul 08826, South Korea}
\newcommand{\instSPU}{Showa Pharmaceutical University, Tokyo 194-8543, Japan}
\newcommand{\instSoochow}{Soochow University, Suzhou 215006, China}
\newcommand{\instSoongsil}{Soongsil University, Seoul 06978, South Korea}
\newcommand{\instLjubljanaJSI}{J. Stefan Institute, 1000 Ljubljana, Slovenia}
\newcommand{\instKyiv}{Taras Shevchenko National Univ. of Kiev, Kiev, Ukraine}
\newcommand{\instTata}{Tata Institute of Fundamental Research, Mumbai 400005, India}
\newcommand{\instTUM}{Department of Physics, Technische Universit\"{a}t M\"{u}nchen, 85748 Garching, Germany}
\newcommand{\instTelAviv}{Tel Aviv University, School of Physics and Astronomy, Tel Aviv, 69978, Israel}
\newcommand{\instToho}{Toho University, Funabashi 274-8510, Japan}
\newcommand{\instTohoku}{Department of Physics, Tohoku University, Sendai 980-8578, Japan}
\newcommand{\instTitech}{Tokyo Institute of Technology, Tokyo 152-8550, Japan}
\newcommand{\instTokyoMetropolitan}{Tokyo Metropolitan University, Tokyo 192-0397, Japan}
\newcommand{\instUAS}{Universidad Autonoma de Sinaloa, Sinaloa 80000, Mexico}
\newcommand{\instNapoliUNIV}{Dipartimento di Scienze Fisiche, Universit\`{a} di Napoli Federico II, I-80126 Napoli, Italy}
\newcommand{\instPadovaUNIV}{Dipartimento di Fisica e Astronomia, Universit\`{a} di Padova, I-35131 Padova, Italy}
\newcommand{\instPerugiaUNIV}{Dipartimento di Fisica, Universit\`{a} di Perugia, I-06123 Perugia, Italy}
\newcommand{\instPisaUNIV}{Dipartimento di Fisica, Universit\`{a} di Pisa, I-56127 Pisa, Italy}
\newcommand{\instRomaTreUNIV}{Dipartimento di Matematica e Fisica, Universit\`{a} di Roma Tre, I-00146 Roma, Italy}
\newcommand{\instTorinoUNIV}{Dipartimento di Fisica, Universit\`{a} di Torino, I-10125 Torino, Italy}
\newcommand{\instTriesteUNIV}{Dipartimento di Fisica, Universit\`{a} di Trieste, I-34127 Trieste, Italy}
\newcommand{\instMontreal}{Universit\'{e} de Montr\'{e}al, Physique des Particules, Montr\'{e}al, Qu\'{e}bec, H3C 3J7, Canada}
\newcommand{\instIJCLab}{Universit\'{e} Paris-Saclay, CNRS/IN2P3, IJCLab, 91405 Orsay, France}
\newcommand{\instIPHC}{Universit\'{e} de Strasbourg, CNRS, IPHC, UMR 7178, 67037 Strasbourg, France}
\newcommand{\instAdelaide}{Department of Physics, University of Adelaide, Adelaide, South Australia 5005, Australia}
\newcommand{\instBonn}{University of Bonn, 53115 Bonn, Germany}
\newcommand{\instUBC}{University of British Columbia, Vancouver, British Columbia, V6T 1Z1, Canada}
\newcommand{\instCincinnati}{University of Cincinnati, Cincinnati, Ohio 45221, U.S.A.}
\newcommand{\instFlorida}{University of Florida, Gainesville, Florida 32611, U.S.A.}
\newcommand{\instHawaii}{University of Hawaii, Honolulu, Hawaii 96822, U.S.A.}
\newcommand{\instHeidelberg}{University of Heidelberg, 68131 Mannheim, Germany}
\newcommand{\instLjubljanaUniLJ}{Faculty of Mathematics and Physics, University of Ljubljana, 1000 Ljubljana, Slovenia}
\newcommand{\instLouisville}{University of Louisville, Louisville, Kentucky 40292, U.S.A.}
\newcommand{\instMalaya}{National Centre for Particle Physics, University Malaya, 50603 Kuala Lumpur, Malaysia}
\newcommand{\instLjubljanaUM}{Faculty of Chemistry and Chemical Engineering, University of Maribor, 2000 Maribor, Slovenia}
\newcommand{\instMelbourne}{School of Physics, University of Melbourne, Victoria 3010, Australia}
\newcommand{\instMississippi}{University of Mississippi, University, Mississippi 38677, U.S.A.}
\newcommand{\instUOM}{University of Miyazaki, Miyazaki 889-2192, Japan}
\newcommand{\instPittsburgh}{University of Pittsburgh, Pittsburgh, Pennsylvania 15260, U.S.A.}
\newcommand{\instUSTC}{University of Science and Technology of China, Hefei 230026, China}
\newcommand{\instSAlabama}{University of South Alabama, Mobile, Alabama 36688, U.S.A.}
\newcommand{\instSCarolina}{University of South Carolina, Columbia, South Carolina 29208, U.S.A.}
\newcommand{\instSydney}{School of Physics, University of Sydney, New South Wales 2006, Australia}
\newcommand{\instUTokyo}{Department of Physics, University of Tokyo, Tokyo 113-0033, Japan}
\newcommand{\instEri}{Earthquake Research Institute, University of Tokyo, Tokyo 113-0032, Japan}
\newcommand{\instIPMU}{Kavli Institute for the Physics and Mathematics of the Universe (WPI), University of Tokyo, Kashiwa 277-8583, Japan}
\newcommand{\instVictoria}{University of Victoria, Victoria, British Columbia, V8W 3P6, Canada}
\newcommand{\instVPI}{Virginia Polytechnic Institute and State University, Blacksburg, Virginia 24061, U.S.A.}
\newcommand{\instWayneState}{Wayne State University, Detroit, Michigan 48202, U.S.A.}
\newcommand{\instYamagata}{Yamagata University, Yamagata 990-8560, Japan}
\newcommand{\instYonsei}{Yonsei University, Seoul 03722, South Korea}
%%%\newcommand{\instZZU}{Zhengzhou University, Zhengzhou 450001, China}
%%%\affiliation{\instSinica}
\affiliation{\instCPPM}
\affiliation{\instYerevan}
\affiliation{\instBeihang}
%%%\affiliation{\instBUAP}
\affiliation{\instBNL}
\affiliation{\instBINP}
\affiliation{\instCMU}
\affiliation{\instCinvestavIPN}
\affiliation{\instPrague}
\affiliation{\instChiangMai}
\affiliation{\instChiba}
\affiliation{\instChonnam}
%%%\affiliation{\instChula}
\affiliation{\instConacyt}
\affiliation{\instDESY}
\affiliation{\instDuke}
\affiliation{\instITAR}
\affiliation{\instRomaENEA}
%%%\affiliation{\instJuelich}
\affiliation{\instFuJen}
\affiliation{\instFudan}
\affiliation{\instGoettingen}
\affiliation{\instGifu}
\affiliation{\instSOKENDAI}
\affiliation{\instGyeongsang}
\affiliation{\instHanyang}
\affiliation{\instKEK}
\affiliation{\instJPARC}
\affiliation{\instHiroshima}
%%%\affiliation{\instHUNNU}
\affiliation{\instFrascati}
\affiliation{\instNapoliINFN}
\affiliation{\instPadovaINFN}
\affiliation{\instPerugiaINFN}
\affiliation{\instPisaINFN}
\affiliation{\instRomaINFN}
\affiliation{\instRomaTreINFN}
\affiliation{\instTorinoINFN}
\affiliation{\instTriesteINFN}
\affiliation{\instIISER}
\affiliation{\instIITBhubaneswar}
\affiliation{\instIITGuwahati}
\affiliation{\instIITHyderabad}
\affiliation{\instIITMadras}
\affiliation{\instIndiana}
\affiliation{\instIHEPRussia}
\affiliation{\instHEPHYVienna}
\affiliation{\instIHEPChina}
%%%\affiliation{\instChennai}
\affiliation{\instIPP}
\affiliation{\instIOP}
\affiliation{\instIFIC}
\affiliation{\instISU}
\affiliation{\instJAEA}
\affiliation{\instMainz}
\affiliation{\instGiessen}
\affiliation{\instKarlsruhe}
%%%\affiliation{\instKennesaw}
\affiliation{\instKitasato}
\affiliation{\instKISTI}
\affiliation{\instKoreaUnivKU}
\affiliation{\instKSU}
%%%\affiliation{\instKyotoU}
\affiliation{\instKyungpook}
\affiliation{\instLPI}
\affiliation{\instLNNU}
\affiliation{\instLMU}
\affiliation{\instLuther}
\affiliation{\instMNITJaipur}
\affiliation{\instMPP}
\affiliation{\instMPGHLL}
\affiliation{\instMcGill}
%%%\affiliation{\instMETU}
\affiliation{\instMEPhI}
\affiliation{\instNagoya}
\affiliation{\instNagoyaIAR}
\affiliation{\instNagoyaKMI}
%%%\affiliation{\instNNU}
\affiliation{\instNaraWu}
%%%\affiliation{\instUNAM}
\affiliation{\instHSE}
\affiliation{\instNTUTaiwan}
\affiliation{\instNUUTaiwan}
\affiliation{\instKrakow}
\affiliation{\instNiigata}
\affiliation{\instNSU}
\affiliation{\instOkinawa}
\affiliation{\instOsakaCity}
\affiliation{\instRCNP}
\affiliation{\instPNNL}
\affiliation{\instPanjab}
%%%\affiliation{\instPeking}
\affiliation{\instPanjabPAU}
\affiliation{\instRIKENMSL}
%%%\affiliation{\instRIKEN}
\affiliation{\instXavier}
\affiliation{\instSeoul}
%%%\affiliation{\instShandong}
\affiliation{\instSPU}
\affiliation{\instSoochow}
\affiliation{\instSoongsil}
\affiliation{\instLjubljanaJSI}
\affiliation{\instKyiv}
\affiliation{\instTata}
\affiliation{\instTUM}
%%%\affiliation{\instECUTUM}
\affiliation{\instTelAviv}
\affiliation{\instToho}
\affiliation{\instTohoku}
\affiliation{\instTitech}
\affiliation{\instTokyoMetropolitan}
\affiliation{\instUAS}
%%%\affiliation{\instNapoliUNIVA}
\affiliation{\instNapoliUNIV}
\affiliation{\instPadovaUNIV}
\affiliation{\instPerugiaUNIV}
\affiliation{\instPisaUNIV}
%%%\affiliation{\instRomaUNIV}
\affiliation{\instRomaTreUNIV}
\affiliation{\instTorinoUNIV}
\affiliation{\instTriesteUNIV}
\affiliation{\instMontreal}
\affiliation{\instIJCLab}
\affiliation{\instIPHC}
\affiliation{\instAdelaide}
\affiliation{\instBonn}
\affiliation{\instUBC}
\affiliation{\instCincinnati}
\affiliation{\instFlorida}
%%%\affiliation{\instHamburg}
\affiliation{\instHawaii}
\affiliation{\instHeidelberg}
\affiliation{\instLjubljanaUniLJ}
\affiliation{\instLouisville}
\affiliation{\instMalaya}
\affiliation{\instLjubljanaUM}
\affiliation{\instMelbourne}
\affiliation{\instMississippi}
\affiliation{\instUOM}
%%%\affiliation{\instNovaGorica}
\affiliation{\instPittsburgh}
\affiliation{\instUSTC}
\affiliation{\instSAlabama}
\affiliation{\instSCarolina}
\affiliation{\instSydney}
%%%\affiliation{\instTabuk}
\affiliation{\instUTokyo}
\affiliation{\instEri}
\affiliation{\instIPMU}
\affiliation{\instVictoria}
\affiliation{\instVPI}
\affiliation{\instWayneState}
\affiliation{\instYamagata}
\affiliation{\instYonsei}
%%%\affiliation{\instZZU}
  \author{F.~Abudin{\'e}n}\affiliation{\instTriesteINFN} % 2250
  \author{I.~Adachi}\affiliation{\instKEK}\affiliation{\instSOKENDAI} % 2590
  \author{R.~Adak}\affiliation{\instFudan} % 6743
  \author{K.~Adamczyk}\affiliation{\instKrakow} % 2239
  \author{L.~Aggarwal}\affiliation{\instPanjab} % 10083
  \author{P.~Ahlburg}\affiliation{\instBonn} % 2367
  \author{H.~Ahmed}\affiliation{\instXavier} % 11323
  \author{J.~K.~Ahn}\affiliation{\instKoreaUnivKU} % 7423
  \author{H.~Aihara}\affiliation{\instUTokyo} % 2223
  \author{N.~Akopov}\affiliation{\instYerevan} % 9443
% ...
% 4983
\collaboration{Belle II Collaboration}

%% file: acknowledgements.tex
We thank the SuperKEKB group for the excellent operation of the
accelerator; the KEK cryogenics group for the efficient
operation of the solenoid; the KEK computer group for
on-site computing support; and the raw-data centers at
BNL, DESY, GridKa, IN2P3, and INFN for off-site computing support.
This work was supported by the following funding sources:
%Armenia
Science Committee of the Republic of Armenia Grant No. 20TTCG-1C010;
%Australia
Australian Research Council and research grant Nos.
DP180102629, 
DP170102389, 
DP170102204, 
DP150103061, 
FT130100303, 
FT130100018,
and
FT120100745;
%Austria
Austrian Federal Ministry of Education, Science and Research,
Austrian Science Fund No. P 31361-N36, and
Horizon 2020 ERC Starting Grant no. 947006 ``InterLeptons''; 
%Canada
Natural Sciences and Engineering Research Council of Canada, Compute Canada and CANARIE;
%China
Chinese Academy of Sciences and research grant No. QYZDJ-SSW-SLH011,
National Natural Science Foundation of China and research grant Nos.
11521505,
11575017,
11675166,
11761141009,
11705209,
and
11975076,
LiaoNing Revitalization Talents Program under contract No. XLYC1807135,
Shanghai Municipal Science and Technology Committee under contract No. 19ZR1403000,
Shanghai Pujiang Program under Grant No. 18PJ1401000,
and the CAS Center for Excellence in Particle Physics (CCEPP);
%Czech Republic
the Ministry of Education, Youth and Sports of the Czech Republic under Contract No.~LTT17020 and 
Charles University grants SVV 260448 and GAUK 404316;
%EU
European Research Council, 7th Framework PIEF-GA-2013-622527, 
Horizon 2020 ERC-Advanced Grants No. 267104 and 884719,
Horizon 2020 ERC-Consolidator Grant No. 819127,
Horizon 2020 Marie Sklodowska-Curie grant agreement No. 700525 `NIOBE,' 
and
Horizon 2020 Marie Sklodowska-Curie RISE project JENNIFER2 grant agreement No. 822070 (European grants);
%France
L'Institut National de Physique Nucl\'{e}aire et de Physique des Particules (IN2P3) du CNRS (France);
%Germany
BMBF, DFG, HGF, MPG, and AvH Foundation (Germany);
%India
Department of Atomic Energy under Project Identification No. RTI 4002 and Department of Science and Technology (India);
%Israel
Israel Science Foundation grant No. 2476/17,
United States-Israel Binational Science Foundation grant No. 2016113, and
Israel Ministry of Science grant No. 3-16543;
%Italy
Istituto Nazionale di Fisica Nucleare and the research grants BELLE2;
%Japan
Japan Society for the Promotion of Science,  Grant-in-Aid for Scientific Research grant Nos.
16H03968, 
16H03993, 
16H06492,
16K05323, 
17H01133, 
17H05405, 
18K03621, 
18H03710, 
18H05226,
19H00682, % Niigata
26220706,
and
26400255,
the National Institute of Informatics, and Science Information NETwork 5 (SINET5), 
and
the Ministry of Education, Culture, Sports, Science, and Technology (MEXT) of Japan;  
%Korea
National Research Foundation (NRF) of Korea Grant Nos.
2016R1\-D1A1B\-01010135,
2016R1\-D1A1B\-02012900,
2018R1\-A2B\-3003643,
2018R1\-A6A1A\-06024970,
2018R1\-D1A1B\-07047294,
2019K1\-A3A7A\-09033840,
and
2019R1\-I1A3A\-01058933,
Radiation Science Research Institute,
Foreign Large-size Research Facility Application Supporting project,
the Global Science Experimental Data Hub Center of the Korea Institute of Science and Technology Information
and
KREONET/GLORIAD;
%Malaysia
Universiti Malaya RU grant, Akademi Sains Malaysia and Ministry of Education Malaysia;
%Mexico
% CINVESTAV-IPN, UNAM, UAS, BUAP and CONACYT are funded under
Frontiers of Science Program contracts
FOINS-296,
CB-221329,
CB-236394,
CB-254409,
and
CB-180023, and SEP-CINVESTAV research grant 237 (Mexico);
%Poland
the Polish Ministry of Science and Higher Education and the National Science Center;
%Russia
the Ministry of Science and Higher Education of the Russian Federation,
Agreement 14.W03.31.0026, and
the HSE University Basic Research Program, Moscow;
%Saudi Arabia
University of Tabuk research grants
S-0256-1438 and S-0280-1439 (Saudi Arabia);
%Slovenia
Slovenian Research Agency and research grant Nos.
J1-9124
and
P1-0135; 
%Spain
Agencia Estatal de Investigacion, Spain grant Nos.
FPA2014-55613-P
and
FPA2017-84445-P,
and
CIDEGENT/2018/020 of Generalitat Valenciana;
%Taiwan
Ministry of Science and Technology and research grant Nos.
MOST106-2112-M-002-005-MY3
and
MOST107-2119-M-002-035-MY3, 
and the Ministry of Education (Taiwan);
%Thailand
Thailand Center of Excellence in Physics;
%Turkey
TUBITAK ULAKBIM (Turkey);
%Ukraine
National Research Foundation of Ukraine, project No. 2020.02/0257,
and
Ministry of Education and Science of Ukraine;
%USA
the US National Science Foundation and research grant Nos.
PHY-1807007 % Luther
and
PHY-1913789, % Indiana CEEM
and the US Department of Energy and research grant Nos.
DE-AC06-76RLO1830, % PNNL
DE-SC0007983, % Wayne State
DE-SC0009824, % Florida
DE-SC0009973, % VPI
DE-SC0010073, % South Carolina
DE-SC0010118, % Carnegie Mellon
DE-SC0010504, % Hawaii
DE-SC0011784, % Cincinnati
DE-SC0012704, % BNL
DE-SC0021274; % Mississippi
%last group
and
%Vietnam
the Vietnam Academy of Science and Technology (VAST) under grant DL0000.05/21-23.

%% file: conf.bbl
\begin{thebibliography}{21}
\bibitem [{cc()}]{cc}%
  \BibitemOpen
  \href@noop {} {}\bibinfo {note} {Charged conjugation is implied throughout
  this analysis.}\BibitemShut {Stop}%
\end{thebibliography}
